\documentclass[10pt,twocolumn,english,manuscript]{revtex4}
\usepackage[T1]{fontenc}
\usepackage[latin9]{inputenc}
\usepackage{amsmath}
\usepackage{amssymb}
\usepackage{graphicx}
\usepackage{esint}

\makeatletter
\@ifundefined{textcolor}{}
{%
 \definecolor{BLACK}{gray}{0}
 \definecolor{WHITE}{gray}{1}
 \definecolor{RED}{rgb}{1,0,0}
 \definecolor{GREEN}{rgb}{0,1,0}
 \definecolor{BLUE}{rgb}{0,0,1}
 \definecolor{CYAN}{cmyk}{1,0,0,0}
 \definecolor{MAGENTA}{cmyk}{0,1,0,0}
 \definecolor{YELLOW}{cmyk}{0,0,1,0}
}

\usepackage{babel}

\usepackage{babel}

\makeatother

\usepackage{babel}
\begin{document}

\title{On the effects of an impurity in an Ising-$XXZ$ diamond chain on
the thermal entanglement, on the quantum coherence and on the quantum
teleportation}

\author{Marcos Freitas$^{1}$, Cleverson Filgueiras$^{1}$ and Moises Rojas$^{1}$}

\affiliation{$^{1}$Departamento de Física, Universidade Federal de Lavras, 37200-000,
Lavras-MG, Brasil}
\begin{abstract}
The effects of an impurity plaquette on the thermal quantum correlations
measurement by the concurrence, on the quantum coherence quantified
by the recently proposed $l_{1}$-norm of coherence and on the quantum
teleportation in a Ising-$XXZ$ diamond chain are discussed. Such
an impurity is formed by the $XXZ$ interaction between the interstitial
Heisenberg dimers and the nearest-neighbor Ising coupling between
the nodal and interstitial spins. All the interaction parameters are
different from those of the rest of the chain. By tailoring them,
the quantum entanglement and quantum coherence can be controlled and
tuned. Therefore, the quantum resources -thermal entanglement and
quantum coherence- of the model exhibit a clear performance improvement
in comparison to the original model without impurities. We also demonstrate
that the quantum teleportation can be tuned by its inclusion. The
thermal teleportation is modified in significant way as well, and
a strong increase in average fidelity is observed. We furnish the
exact solution by the use of the transfer-matrix method. 
\end{abstract}
\maketitle

\section{Introduction}

The quantum resource theories \cite{st,st-1} play a central role
in the quantum information processing. In particular, quantum coherence
and entanglement are resources for quantum technological applications
i.e., quantum communication and quantum computation \cite{bra,Bene,amico}.
Recently, the role of the quantum coherence on Heisenberg spin models
has been considered \cite{rada,fan,wei}. Furthermore, recent research
reveals that these resources have a close connection with each other\cite{adesso}.
On the other hand, quantum entanglement is one of the most fascinating
features of the quantum theory, and it has been regarded as an essential
physical resource for quantum computation and quantum information.
The Heisenberg chain is one of the simplest quantum system which exhibits
entanglement. For this reason, the Heisenberg spin models have been
extensively studied in condensed matter systems\cite{kam}. Also,
many schemes of teleportation via thermal entanglement states have
been reported\cite{yeo}. In the context of the spin-1/2 Heisenberg
model with a diamond chain structure, a novel class of the simplified
versions of the so-called Ising-Heisenberg diamond chain was introduced
in Ref. \cite{strec}. The various thermodynamic properties of it
have been extensively investigated\cite{rojas}. Recently, the thermal
quantum entanglement in some exactly solvable Ising-Heisenberg diamond
chains have been extensively analyzed and discussed \cite{moi,cheng,rojas-1,rojas-2}.
More recently, Rojas $et\,al$ \cite{moi-1} discussed the entangled
state teleportation through a couple of quantum channels composed
of $XXZ$ dimers in an Ising-$XXZ$ diamond chain.

Impurities play an important role in solid state physics \cite{falk}.
Even a small defect may changes the physical properties of the quantum
system. In recent years, the study of the spin chains with impurities
has attracted much attention \cite{fuku,xuchu}, including the various
kinds of the spin chains with a magnetic impurity\cite{fu}. In Ref.
\cite{rojas-3}, the tuning of the thermal entanglement in a Ising-$XXZ$
diamond chain with two impurities was addressed.

Motivated by these mentioned developments, the present work is addressed
to a detailed investigation on the influence of an impurity plaquette
inserted in an Ising-$XXZ$ diamond chain. Such an impurity spin is
defined by a local change in the nearest-neighbor couplings. We will
focus on the analysis of the thermal entanglement and on the quantum
coherence in this impurity embedded environment. It is shown that
the impurities parameters can generate a significant enhancement on
the entanglement and on the quantum coherence. Besides, we study the
teleportation of an unknown state using a couple of impure Heisenberg
dimers embedded in an Ising-$XXZ$ diamond chain in thermal equilibrium
as a quantum channel. The effects due to it as well as those due to
the parameters of a Heisenberg interaction, an Ising interaction and
magnetic fields in the fidelity and average fidelity are obtained
analytically.

The organization of this article is as follows. In Sec. II we introduce
the Ising-XXZ model with an impurity. In Sec.III, we obtain the exact
solution of the model via the transfer-matrix approach and its dimer
(two-qubit) reduced density operator. In Sec.IV, we discuss the thermal
entanglement and quantum coherence of the impurity Heisenberg reduced
density operator of the model. In Sec.V, we study the effects of the
impurity parameters on teleportation scheme. We evaluate the fidelity
and average fidelity. Finally, the concluding remarks are given in
Sec. VI. 

\section{The Model}

\begin{figure}
\includegraphics[scale=0.44]{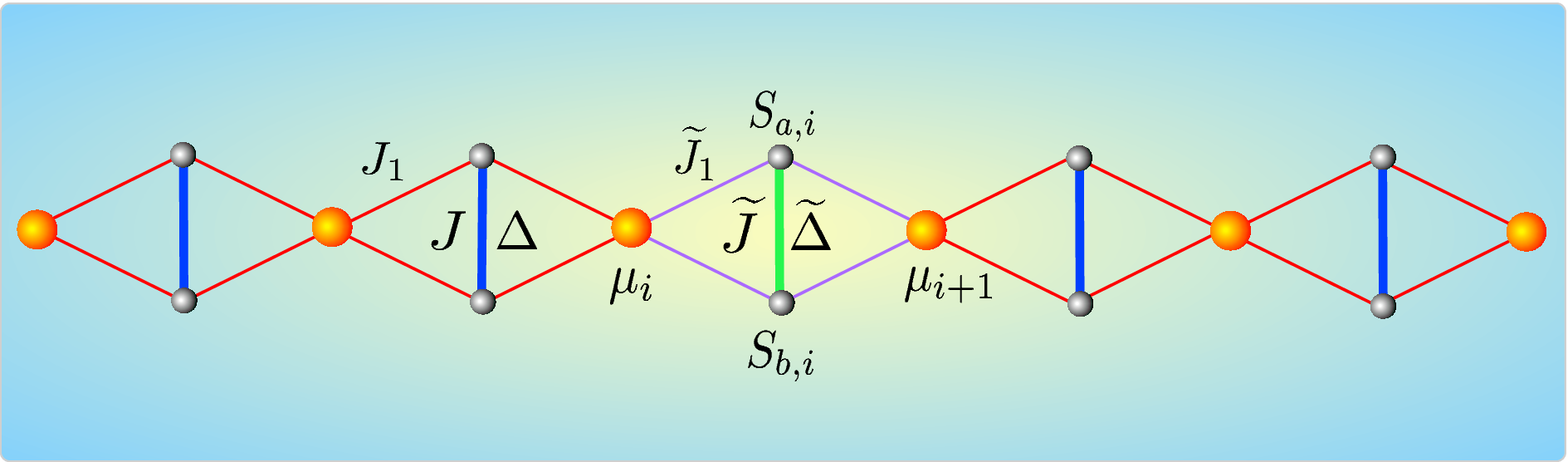} \caption{\label{fig:diamond}(Color online) A schematic representation of an
Ising-$XXZ$ diamond chain with one impurity inserted in the $i$-th
block of the primitive unit cell. }
\end{figure}

In this section, we introduce the hamiltonian of the spin-1/2 Ising-$XXZ$
model on a diamond chain with one plaquette impurity under an external
magnetic field, $h$. The model consists on the interstitial Heisenberg
spins $(S_{a,i},\:S_{b,i})$ and Ising spins $(\sigma_{i},\:\sigma_{i+1})$
located in the nodal site, as shown in Fig. \ref{fig:diamond}. The
total hamiltonian of the model may be written as 
\[
\mathcal{H}=\sum_{i=1}^{N}\mathcal{H}_{i},
\]
where $\mathcal{H}_{i}=\mathcal{H}_{i}^{host}+\mathcal{H}_{i}^{imp},$
and the host hamiltonian $\mathcal{H}_{i}^{host}$ can be expressed
as 
\[
\begin{array}{cl}
\mathcal{H}_{i}^{host}= & J\left(\mathbf{S}_{a,i},\mathbf{S}_{b,i}\right)_{\Delta}+J_{1}\left(S_{a,i}^{z}+S_{b,i}^{z}\right)\left(\mu_{i}+\mu_{i+1}\right)\\
 & -h\left(S_{a,i}^{z}+S_{b,i}^{z}\right)-\frac{h}{2}\left(\mu_{i}+\mu_{i+1}\right).\\
 & \mathrm{for}\:i=1,2,\ldots,r-1,r+1,\ldots,N
\end{array}
\]
On the other hand, the impurity induced hamiltonian $\mathcal{H}_{i}^{imp}$,
is defined by 
\[
\begin{array}{cl}
\mathcal{H}_{i}^{imp}= & \widetilde{J}\left(\mathbf{S}_{a,i},\mathbf{S}_{b,i}\right)_{\widetilde{\Delta}}+\widetilde{J}_{1}\left(S_{a,i}^{z}+S_{b,i}^{z}\right)\left(\mu_{i}+\mu_{i+1}\right)\\
 & -h\left(S_{a,i}^{z}+S_{b,i}^{z}\right)-\frac{h}{2}\left(\mu_{i}+\mu_{i+1}\right),\\
 & \mathrm{for}\:i=r\;,
\end{array}
\]
where the parameters $J$ and $\Delta$ denote the $XXZ$ interaction
within the Heisenberg dimer, the nodal-interstitial spins interaction
are represented by the Ising-type exchanges $J_{1}$, $h$ denotes
the longitudinal magnetic field in the $z$ direction and the impurity
parameters are given by $\widetilde{J}=J\left(1+\alpha\right)$, $\widetilde{\Delta}=\Delta\left(1+\gamma\right)$
and $\widetilde{J}_{1}=J_{1}\left(1+\eta\right)$.

After straightforward calculations, the eigenvalues for the $XXZ$
dimer of the above host hamiltonian $\mathcal{H}_{i}^{host}$ can
be obtained as 
\begin{align*}
\mathcal{E}_{i1,i4}= & \frac{J\Delta}{4}\pm\left(J_{1}\mp\frac{h}{2}\right)(\mu_{i}+\mu_{i+1})\mp h,\\
\mathcal{E}_{i2,i3}= & -\frac{J\Delta}{4}\pm\frac{J_{1}}{2}-\frac{h}{2}(\sigma_{i}+\sigma_{i+1}),
\end{align*}
where their corresponding eigenstates in terms of the standard basis
$\{|00\rangle,|01\rangle,|10\rangle,|11\rangle\}$ are given, respectively,
by 
\begin{align}
|\varphi_{i1}\rangle= & |00\rangle_{i},\\
|\varphi_{i2}\rangle= & \frac{1}{\sqrt{2}}\left(|01\rangle_{i}+|10\rangle_{i}\right),\label{eq:phi2}\\
|\varphi_{i3}\rangle= & \frac{1}{\sqrt{2}}\left(|01\rangle_{i}-|10\rangle_{i}\right),\\
|\varphi_{i4}\rangle= & |11\rangle_{i}\;.
\end{align}
Analogously to the impurity dimer, the eigenvalues of the $\mathcal{H}_{i}^{imp}$
are 
\begin{align*}
\mathcal{\widetilde{E}}_{i1,i4}= & \frac{\widetilde{J}\widetilde{\Delta}}{4}\pm\left(\widetilde{J}_{1}\mp\frac{h}{2}\right)(\mu_{i}+\mu_{i+1})\mp h,\\
\mathcal{\widetilde{E}}_{i2,i3}= & -\frac{\widetilde{J}\widetilde{\Delta}}{4}\pm\frac{\widetilde{J}_{1}}{2}-\frac{h}{2}(\sigma_{i}+\sigma_{i+1}),
\end{align*}
and the corresponding eigenstates are 
\begin{align}
|\widetilde{\varphi}_{i1}\rangle= & |00\rangle_{i},\\
|\widetilde{\varphi}_{i2}\rangle= & \frac{1}{\sqrt{2}}\left(|01\rangle_{i}+|10\rangle_{i}\right),\label{eq:phi2-1}\\
|\widetilde{\varphi}_{i3}\rangle= & \frac{1}{\sqrt{2}}\left(|01\rangle_{i}-|10\rangle_{i}\right),\\
|\widetilde{\varphi}_{i4}\rangle= & |11\rangle_{i}.
\end{align}
Here, $r=i$. 

\section{The partition function and the density operator}

In order to study the thermal entanglement, the quantum coherence
and the quantum teleportation, we first must obtain a partition function
for a diamond chain. This model can be solved exactly through the
transfer-matrix approach \cite{baxter}. In order to summarize this
approach we will define the following operator, as a function of Ising
spin particles $\mu_{i}$ and $\mu_{i+1}$, 
\begin{equation}
\varrho(\mu_{i},\mu_{i+1})=\sum_{i=1}^{4}\mathrm{e}^{-\beta\mathcal{E}_{ij}(\mu_{i},\mu_{i+1})}|\varphi_{ij}\rangle\langle\varphi_{ij}|\;,\label{eq:rho-loc}
\end{equation}
where $\beta=1/k_{B}T$, $k_{B}$ is the Boltzmann's constant and
$T$ is the absolute temperature.

Straightforwardly, we can obtain the Boltzmann factor by tracing out
over the two-qubit operator, 
\begin{equation}
w(\mu_{i},\mu_{i+1})=\mathrm{tr}_{ab}\left(\widetilde{\varrho}(\mu_{i},\mu_{i+1})\right)=\sum_{j=1}^{4}\mathrm{e}^{-\beta\mathcal{E}_{ij}(\mu_{i},\mu_{i+1})}\;.\label{eq:w-def}
\end{equation}
The Boltzmann factor for an impurity is given by 
\[
\widetilde{w}(\mu_{i},\mu_{i+1})=\sum_{j=1}^{4}\mathrm{e}^{-\beta\mathcal{\widetilde{E}}_{ij}(\mu_{i},\mu_{i+1})}\;,
\]
where $i=r$. The Ising-XXZ diamond chain partition function can be
written in terms of the Boltzmann factors, 
\begin{align}
Z_{N}= & \sum_{\{\mu\}}w(\mu_{1},\mu_{2})\ldots w(\mu_{r-1},\mu_{r})\widetilde{w}(\mu_{r},\mu_{r+1})\times\nonumber \\
 & w(\mu_{r+1},\mu_{r+2})\ldots w(\mu_{N},\mu_{1})\;.\label{eq:rho-df-1}
\end{align}
Using the transfer-matrix notation, we can write the partition function
of the diamond chain straightforwardly by $Z_{N}=\mathrm{tr}\left(\widetilde{W}W^{N-1}\right),$
where the transfer-matrix is expressed as 
\begin{equation}
W=\left[\begin{array}{cc}
w(\frac{1}{2},\frac{1}{2}) & w(\frac{1}{2},-\frac{1}{2})\\
w(-\frac{1}{2},\frac{1}{2}) & w(-\frac{1}{2},-\frac{1}{2})
\end{array}\right]\;.\label{eq:W}
\end{equation}
A similar formula can also be derived for the transfer-matrix $\widetilde{W}$
for the impurity case, namely 
\[
\widetilde{W}=\left[\begin{array}{cc}
\widetilde{w}(\frac{1}{2},\frac{1}{2}) & \widetilde{w}(\frac{1}{2},-\frac{1}{2})\\
\widetilde{w}(-\frac{1}{2},\frac{1}{2}) & \widetilde{w}(-\frac{1}{2},-\frac{1}{2})
\end{array}\right]\;.
\]
In it, the transfer matrix elements are denoted by $w_{++}\equiv w(\frac{1}{2},\frac{1}{2})$,
$w_{+-}\equiv w(\frac{1}{2},-\frac{1}{2})$ and $w_{--}\equiv w(-\frac{1}{2},-\frac{1}{2})$.
After performing the diagonalization of the transfer matrix (\ref{eq:W}),
the eigenvalues are found, that is, 
\begin{equation}
\Lambda_{\pm}=\frac{w_{++}+w_{--}\pm Q}{2}\;.
\end{equation}
It was assumed that $Q=\sqrt{\left(w_{++}-w_{--}\right)^{2}+4w_{+-}^{2}}$.
Therefore, the partition function for finite chain under periodic
boundary conditions is given by 
\begin{equation}
Z_{N}=a\Lambda_{+}^{N-1}+d\Lambda_{-}^{N-1}\;,
\end{equation}
where 
\[
\begin{array}{cc}
a= & \frac{4w_{+-}\widetilde{w}_{+-}+\left(w_{++}-w_{--}\right)\left(\widetilde{w}_{++}-\widetilde{w}_{--}\right)+Q\left(\widetilde{w}_{++}+\widetilde{w}_{--}\right)}{2Q},\\
d= & \frac{-4w_{+-}\widetilde{w}_{+-}-\left(w_{++}-w_{--}\right)\left(\widetilde{w}_{++}-\widetilde{w}_{--}\right)+Q\left(\widetilde{w}_{++}+\widetilde{w}_{--}\right)}{2Q}\;.
\end{array}
\]
In the thermodynamic limit, the partition function will be simplified.
Thus we obtain $Z_{N}=a\Lambda_{+}^{N-1}$. Now, we are interested
in the thermal quantum correlations and in the quantum teleportation.
To reach out our goal it is essential to obtain the reduced density
operators $\widetilde{\rho}$ of the dimer Heisenberg impurity. 

\subsection{Two-qubit operator }

In order to calculate the thermal average of the two-qubit operator
corresponding to an impurity, also called reduced two-qubit density
operator, we will use the approach recently studied in Ref. \cite{rojas-3}.
This way, we will define the operator $\widetilde{\varrho}$ for an
impurity as a function of the Ising particles $\mu_{r}$ and $\mu_{r+1}$,
that is, 
\begin{equation}
\widetilde{\varrho}(\mu_{r},\mu_{r+1})=\left[\begin{array}{cccc}
\widetilde{\varrho}_{1,1} & 0 & 0 & 0\\
0 & \widetilde{\varrho}_{2,2} & \widetilde{\varrho}_{2,3} & 0\\
0 & \widetilde{\varrho}_{2,3} & \widetilde{\varrho}_{2,2} & 0\\
0 & 0 & 0 & \widetilde{\varrho}_{4,4}
\end{array}\right]\;,
\end{equation}
where the elements of the two-qubit operator are given by 
\begin{align*}
\widetilde{\varrho}_{1,1}(\mu_{r},\mu_{r+1})= & \mathrm{e}^{-\beta\mathcal{\widetilde{E}}_{r1}(\mu_{r},\mu_{r+1})},\\
\widetilde{\varrho}_{2,2}(\mu_{r},\mu_{r+1})= & \frac{1}{2}\left(\mathrm{e}^{-\beta\mathcal{\widetilde{E}}_{r2}(\mu_{r},\mu_{r+1})}+\mathrm{e}^{-\beta\mathcal{\widetilde{E}}_{r3}(\mu_{r},\mu_{r+1})}\right),\\
\widetilde{\varrho}_{2,3}(\mu_{r},\mu_{r+1})= & \frac{1}{2}\left(\mathrm{e}^{-\beta\mathcal{\widetilde{E}}_{r2}(\mu_{r},\mu_{r+1})}-\mathrm{e}^{-\beta\mathcal{\widetilde{E}}_{r3}(\mu_{r},\mu_{r+1})}\right),\\
\widetilde{\varrho}_{4,4}(\mu_{r},\mu_{r+1})= & \mathrm{e}^{-\beta\mathcal{\widetilde{E}}_{r4}(\mu_{r},\mu_{r+1})}\;.
\end{align*}
The thermal average for each two-qubit Heisenberg operator will be
used to construct the reduced density operator. 

\subsection{The reduced density operator for the impurity}

The elements of the reduced density operator, $\widetilde{\rho}_{k,l}$,
for an impurity localized in the $r$th block (unit cell), can be
defined by 
\begin{align}
\widetilde{\rho}_{k,l}= & \frac{1}{Z_{N}}\sum_{\{\mu\}}w(\mu_{1},\mu_{2})\ldots w(\mu_{r-1},\mu_{r})\widetilde{\varrho}_{k,l}(\mu_{r},\mu_{r+1})\times\nonumber \\
 & w(\mu_{r+1},\mu_{r+2})\ldots w(\mu_{N},\mu_{1})\;.\label{eq:rho-df}
\end{align}
Using the transfer-matrix approach, $\widetilde{\rho}_{k,l}$ can
be alternatively rewritten as 
\begin{equation}
\tilde{\rho}_{k,l}=\frac{1}{Z_{N}}\mathrm{tr}\left(W^{r-1}\widetilde{P}_{k,l}W^{N-r}\right)=\frac{1}{Z_{N}}\mathrm{tr}\left(\widetilde{P}_{k,l}W^{N-1}\right)\;,
\end{equation}
where 
\begin{equation}
\widetilde{P}_{k,l}=\left[\begin{array}{cc}
\widetilde{\varrho}_{k,l}(\tfrac{1}{2},\tfrac{1}{2}) & \widetilde{\varrho}_{k,l}(\tfrac{1}{2},-\tfrac{1}{2})\\
\widetilde{\varrho}_{k,l}(-\tfrac{1}{2},\tfrac{1}{2}) & \widetilde{\varrho}_{k,l}(-\tfrac{1}{2},-\tfrac{1}{2})
\end{array}\right],
\end{equation}
and $\widetilde{\varrho}_{k,l}(++)\equiv\widetilde{\varrho}_{k,l}(\frac{1}{2},\frac{1}{2})$,
$\widetilde{\varrho}_{k,l}(+-)\equiv\widetilde{\varrho}_{k,l}(\frac{1}{2},-\frac{1}{2})$,
$\widetilde{\varrho}_{k,l}(-+)\equiv\widetilde{\varrho}_{k,l}(-\frac{1}{2},\frac{1}{2})$,
$\widetilde{\varrho}_{k,l}(--)\equiv\widetilde{\varrho}_{k,l}(-\frac{1}{2},-\frac{1}{2})$.
The unitary transformation that diagonalizes the transfer matrix $W$
is determined by $U$, which is given by 
\begin{align}
U=\left[\begin{array}{cc}
\Lambda_{+}-w_{--} & \Lambda_{-}-w_{--}\\
w_{+-} & w_{+-}
\end{array}\right],
\end{align}
and 
\begin{align}
U^{-1}=\left[\begin{array}{cc}
\frac{1}{Q} & -\frac{\Lambda_{-}-w_{--}}{Qw_{+-}}\\
-\frac{1}{Q} & \frac{\Lambda_{+}-w_{--}}{Qw_{+-}}
\end{array}\right]\;.
\end{align}
Finally, the individual matrix elements reduced density operator on
the impurity defined in Eq.(\ref{eq:rho-df}) must be expressed by
\begin{equation}
\widetilde{\rho}_{k,l}=\tfrac{\mathrm{tr}\left(U^{-1}\widetilde{P}_{k,l}U\left[\begin{smallmatrix}\Lambda_{+}^{N-1} & 0\\
0 & \Lambda_{-}^{N-1}
\end{smallmatrix}\right]\right)}{a\Lambda_{+}^{N-1}+d\Lambda_{-}^{N-1}}\;.
\end{equation}
This result is valid for arbitrary number $N$ of cells in the diamond
chain. In the thermodynamic limit, ($N\rightarrow\infty$), the reduced
density operator elements, after some algebraic manipulation, becomes
\begin{align*}
\widetilde{\rho}_{k,l}= & \frac{\mathcal{A}_{k,l}+\mathcal{B}_{k,l}}{\mathcal{M}}\;,
\end{align*}
where 
\[
\begin{array}{cl}
\mathcal{A}_{k,l}= & Q\left[\widetilde{\varrho}_{k,l}(++)+\widetilde{\varrho}_{k,l}(--)\right]+4w_{+-}\widetilde{\varrho}_{k,l}(+-),\\
\mathcal{B}_{k,l}= & \left[\widetilde{\varrho}_{k,l}(++)-\widetilde{\varrho}_{k,l}(--)\right]\left(w_{++}-w_{--}\right),\\
\mathcal{M}= & Q\left(\widetilde{w}_{++}+\widetilde{w}_{--}\right)+4w_{+-}\widetilde{w}_{+-}+\\
 & +\left(\widetilde{w}_{++}-\widetilde{w}_{--}\right)\left(w_{++}-w_{--}\right)\;.
\end{array}
\]
All the elements of the reduced density operator due to the impurity
immersed on a diamond chain are 
\begin{equation}
\widetilde{\rho}(T)=\left[\begin{array}{cccc}
\widetilde{\rho}_{1,1} & 0 & 0 & 0\\
0 & \widetilde{\rho}_{2,2} & \widetilde{\rho}_{2,3} & 0\\
0 & \widetilde{\rho}_{2,3} & \widetilde{\rho}_{2,2} & 0\\
0 & 0 & 0 & \widetilde{\rho}_{4,4}
\end{array}\right]\;.\label{eq:rho-imp}
\end{equation}
It is worth to notice that, such an impurity reduced density operator
is the thermal average two-qubit Heisenberg operator, immersed in
the diamond chain and it can be verified that ${\rm tr}(\widetilde{\rho})=1$.

\section{The thermal entanglement and the quantum coherence of the two-qubit
Heisenberg impurity}


In this section, we aim to study the impurity effects of our model
on the thermal entanglement and on the quantum coherence. The quantum
entanglement is a special type of correlation, which only arises in
quantum systems. In order to measure the entanglement of the anisotropic
Heisenberg qubits in the Ising-Heisenberg model on a diamond chain,
we study the concurrence (entanglement) of the two-qubits Heisenberg
(dimer), which interacts with two nodal Ising spins. We use Wootters
concurrence $\mathcal{C}$ to measure the entanglement \cite{hill,woo},
that is, 
\begin{equation}
\mathcal{C}(\rho)=\mathrm{max}\{\sqrt{\lambda_{1}}-\sqrt{\lambda_{2}}-\sqrt{\lambda_{3}}-\sqrt{\lambda_{4}},0\}\;,\label{eq:Concurrence}
\end{equation}
where the parameters $\lambda_{i}$ $\left(i=1,2,3,4\right)$ are
eigenvalues in the decreasing order of the operator $R$, which is
given by 
\begin{equation}
R=\rho\cdot\left(\sigma^{y}\otimes\sigma^{y}\right)\cdot\rho^{*}\cdot\left(\sigma^{y}\otimes\sigma^{y}\right)\;.\label{eq:R}
\end{equation}
$\rho^{*}$ denotes the complex conjugate of matrix $\rho$. In our
model, substituting the Eq. (\ref{eq:rho-imp}) in the Eq. (\ref{eq:R}),
we get the concurrence of two-qubits Heisenberg impurity, 
\begin{equation}
\mathcal{C}(\widetilde{\rho})=2\mathrm{max}\{|\widetilde{\rho}_{2,3}|-\sqrt{\widetilde{\rho}_{1,1}\widetilde{\rho}_{4,4}},0\}\;.\label{eq:conc-def}
\end{equation}
The thermal entanglement in a Ising-$XXZ$ diamond chain was addressed
in Ref. \cite{moi}. In it, the effects on the thermal entanglement
in an exactly solvable Ising-$XXZ$ diamond chain were investigated.
In the present work, we are interested in investigating the effects
on thermal entanglement, the quantum coherence as well as the quantum
teleportation caused by the impure dimer isolated on an Ising-$XXZ$
diamond chain.

On the other hand, the quantum coherence is a useful resource for
the quantum information processing task. Here, we will employ the
$l_{1}$-norm\cite{baum} measure, defined as 
\[
C_{l_{1}}(\rho)=\underset{i\neq j}{\sum}|\rho_{ij}|\;.
\]
The corresponding $l_{1}$-norm of the quantum coherence of the impure
dimer described by the reduced density operator, $\widetilde{\rho}(T)$(Eq.\ref{eq:rho-imp}),
is given by 
\[
C_{l_{1}}=2|\widetilde{\rho}_{2,3}|.
\]

\subsection{Concurrence}

\begin{figure}
\includegraphics[scale=0.4]{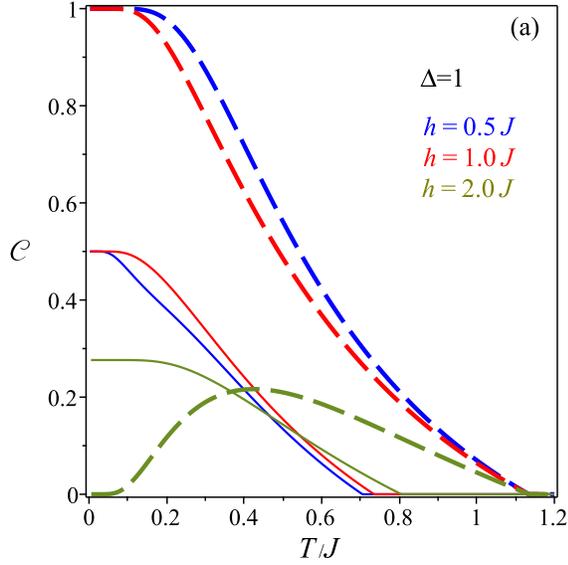}

\includegraphics[scale=0.4]{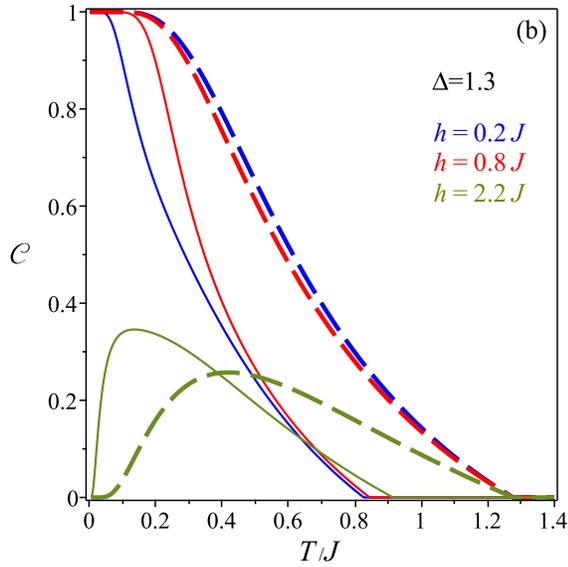}

\caption{\label{fig:CvsT}(Color online) The concurrence $\mathcal{C}$ as
a function of $T/J$, with $J_{1}/J=1$. For the model Ising-$XXZ$
without impurities (solid curve), we have $\alpha=0$, $\gamma=0$
and $\eta=0$. With an impurity(dashed curve), we fixed $\alpha=0$,
$\gamma=0.8$ and $\eta=-0.8$. (a) $\Delta=1.0$ . (b) $\Delta=1.3$. }
\end{figure}

From now on, we will plot curves of the original model with solid
lines, while that for the Ising-$XXZ$ diamond chain with one impurity,
we will use dashed lines. Also, we set the parameters of the impurity
as $\alpha=0$, $\gamma=0.8$ and $\eta=-0.8$. In Fig. \ref{fig:CvsT}(a),
the thermal concurrence, $\mathcal{C}$, as a function of the temperature
$T/J$ is depicted for $\Delta=1$ (Ising-$XXX$ isotropic model)
and for several values of the magnetic field, $h/J$. It is shown
that, for the weak magnetic fields $h/J$, the concurrence in the
dimer with an impurity is maximally entangled, in contrast to the
model without impurity in which it is partially entangled. So, the
behavior of the concurrence is more robust in the presence of an impurity.
For strong magnetic fields, the concurrence is null for low temperatures
in this case. However, with the temperature increasing, we have a
sudden birth of entanglement until the maximum value $\mathcal{C}\approx0.22$
is reached. Then, it completely disappears in the threshold temperature,
$T/J\approx1.12$. It should be noticed that, in this case, the model
without impurity is partially entangled at low temperatures, in contrast
for the case with higher temperatures, where the impurity enhances
the thermal entanglement. In Fig. \ref{fig:CvsT}(b), we plot the
concurrence $\mathcal{C}$ as a function of temperature $T/J$ for
the anisotropic parameter $\Delta=1.3$ and for several values of
the magnetic field. For weak magnetic field and for the temperature
$0\leqslant T/J\lesssim0.2$, the behavior of thermal entanglement
is maximum for both models. When the temperature increases it is possible
to observe the strong influence of the impurity. The entanglement
is more robust and the threshold temperature increases in the model
with the impurity, reaching the value $T/J\approx1.26$. For strong
magnetic fields, the concurrence disappears for the impurity case.
One can observe that, as the temperature increases, for the phenomenon
of the entanglement, a sudden birth occurs and the concurrence reaches
the value $\mathcal{C}\approx0.26$. Then, it suddenly disappears
at $T/J\approx1.26$. Furthermore, it can be seen that the threshold
temperature is improved for various values of magnetic field in the
Ising-$XXZ$ diamond chain with impurity. 


\subsection{The $l_{1}$-norm of coherence}

\begin{figure}
\includegraphics[scale=0.4]{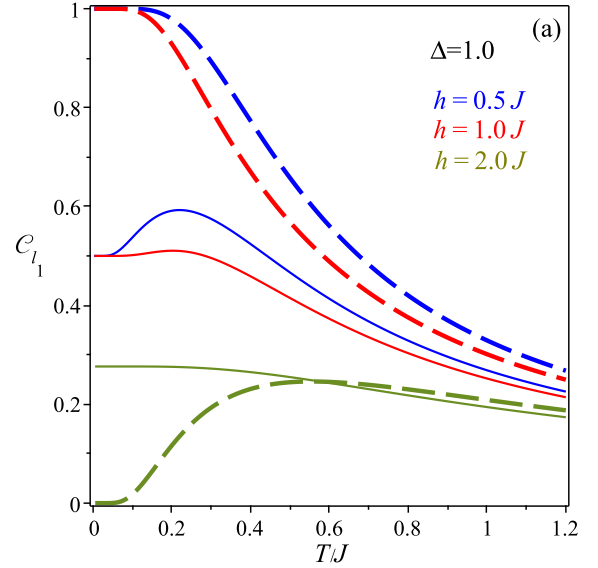}

\includegraphics[scale=0.4]{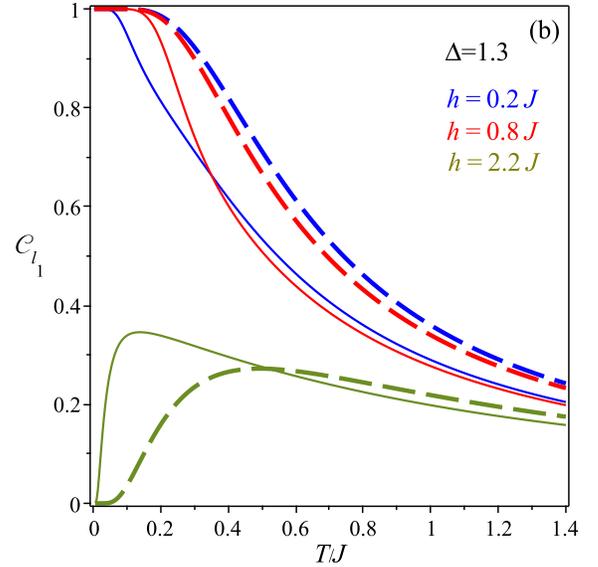}\caption{\label{fig:Cl1}(Color online) The $l_{1}$-norm of coherence $\mathcal{C}_{l_{1}}$
as a function of temperature $T/J$ for $J_{1}/J=1.0$ and different
values of the magnetic field $h$. (a) $\Delta=1.0$. (b) $\Delta=1.3$.
The impurity parameters are set to $\alpha=0$, $\gamma=0.8$ and
$\eta=-0.8$. }
\end{figure}

In order to illustrate the quantum coherence in our model, the $l_{1}$-norm
$\mathcal{C}_{l_{1}}$ versus the temperature $T/J$ is plotted in
Fig. \ref{fig:Cl1} for different values of the magnetic field and
for the anisotropic parameter $\Delta=1$ and $\Delta=1.3$, respectively.
In Fig. \ref{fig:Cl1}(a), we observe a dramatic increase in the $l_{1}$-norm
$\mathcal{C}_{l_{1}}$ for weak magnetic fields, when the impurity
is included in the model. For the strong magnetic field case, $h/J=2.0$,
the impurity diamond chain has a singular behavior in the quantum
coherence. For low temperatures, the quantum coherence is null. Suddenly,
the system increases the coherence, reaching the value $\mathcal{C}_{l_{1}}\approx0.25$
and decreasing monotonically as soon as the $T/J$ increases. In Fig.
\ref{fig:Cl1}(b), the results of $\mathcal{C}_{l_{1}}$ indicate
that, for low temperatures, the quantum coherence is maximum, that
is, $\mathcal{C}_{l_{1}}=1.0$, for both models. However, for the
higher temperature, the quantum coherence with impurity is more robust
for the weak magnetic field case in comparison to that without it.
On the other hand, for strong magnetic fields, the quantum coherence
behavior is very similar to that of the concurrence: at first, it
is null and soon a sudden birth occurs until it reaches the value
$\mathcal{C}_{l_{1}}\approx0.28$. Soon after, it monotonically decreases
with the increasing of the temperature. It is interesting to notice
that both the concurrence and the quantum coherence have the same
behavior at low temperatures, but when the temperature increases,
the quantum coherence happens to predominate over the quantum entanglement(see
Fig. \ref{fig:CvsT} and Fig. \ref{fig:Cl1}). 


\subsection{Threshold temperature}

\begin{figure}
\includegraphics[scale=0.55]{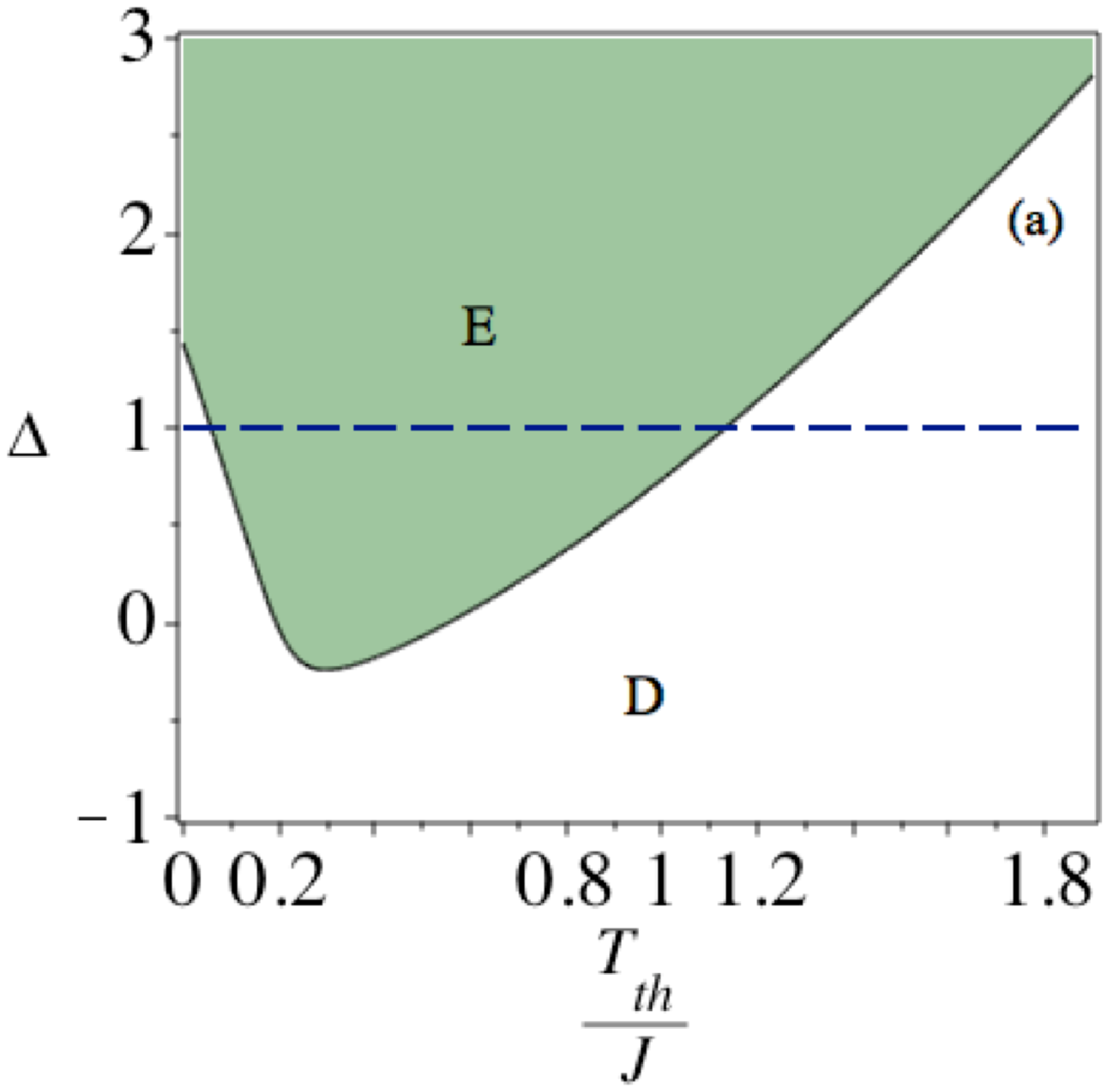} \includegraphics[scale=0.55]{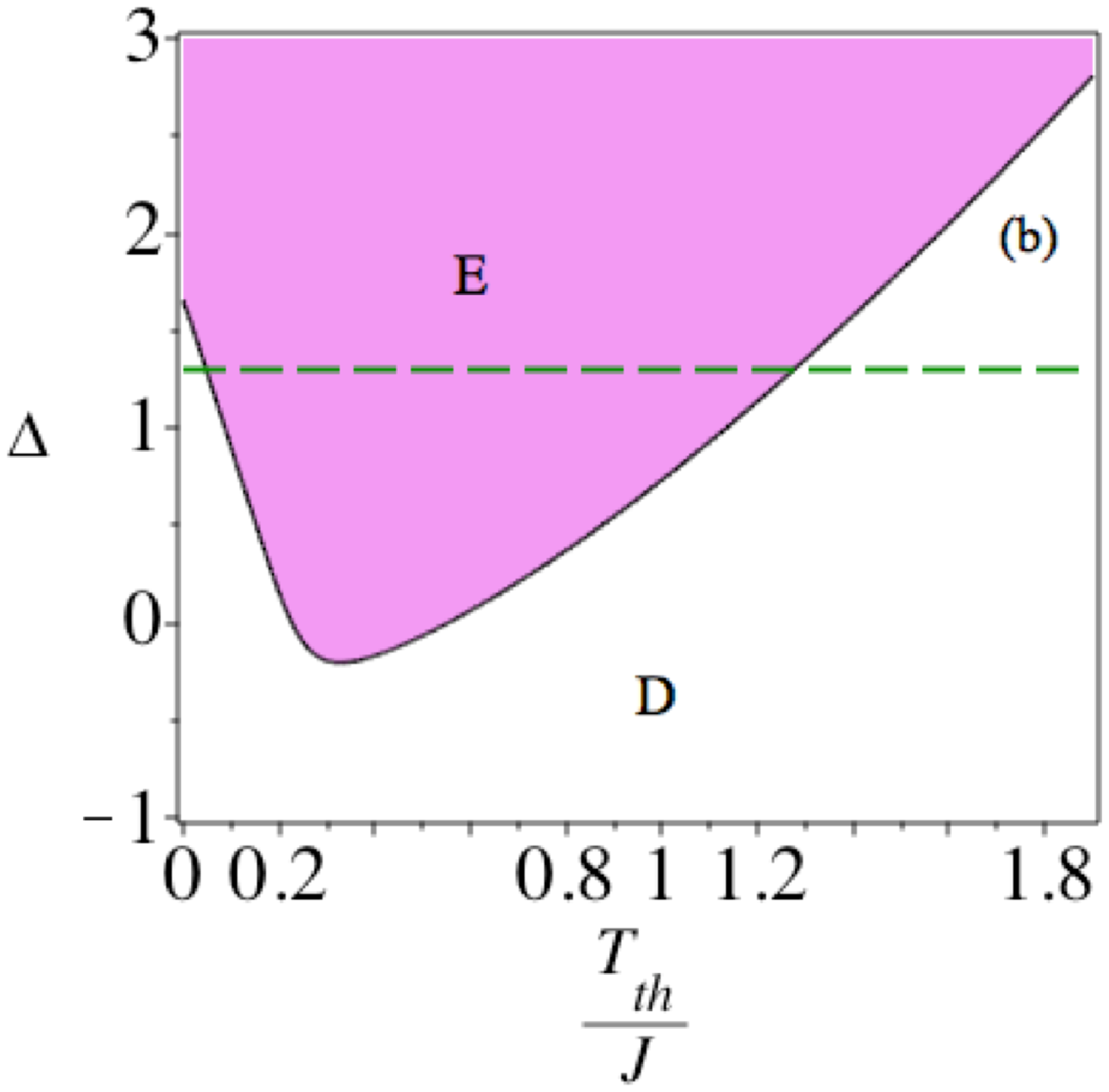}

\caption{\label{fig:diagrama}(Color online) The concurrence $\mathcal{C}$
depending on the anisotropy parameter and threshold temperature, when
$J_{1}/J=1$ and $\alpha=0$, $\gamma=0.8$ and $\eta=-0.8$. (a)
$h/J=2.0$ and $\Delta=1.0$ (blue dashed line). (b) $h/J=2.2$ and
$\Delta=1.3$ (green dashed line).}
\end{figure}

We now investigated the effects of the temperature on the behavior
of the thermal concurrence and on the quantum coherence with the presence
of a strong external magnetic field. Figures \ref{fig:CvsT} and \ref{fig:Cl1}
show an unusual behavior in the concurrence and in the quantum coherence
($l_{1}$-norm) for strong magnetic fields ($h/J=2.0$ and $h/J=2.2$).
In Fig. \ref{fig:diagrama}, we display the phase diagram of the entangled
region (E) and the disentagled region (D), as a function of both the
anisotropy parameter $\Delta$ and the threshold temperature $T_{th}/J$
for some values of the magnetic field and the fixed value $J_{1}/J=1$.
Also, we set the parameters of the impurity as $\alpha=0$, $\gamma=0.8$
and $\eta=-0.8$. Here, the threshold temperature, delimits the regime
of the entangled spins (finite concurrence) and the disentangled spins
(vanishing concurrence). The entangled region (E) is the entangled
state in the quantum ferrimagnetic phase, which was denoted by $ENQ$,
while the disentangled region (D) is the unentangled ferromagnetic
state which is denoted by $UFM$ (see Ref. \cite{moi}). It is observed
from Fig. \ref{fig:diagrama} (a), for strong magnetic field ($h/J=2.0$),
the re-entrant behavior of the thermal entanglement when the anisotropy
parameter ($\Delta=1$) is sufficiently close but slightly below the
ground-state boundary between the $ENQ$ and $UFM$ phases (see Ref.
\cite{moi}). Under this condition, the concurrence/coherence starts
from zero (see blue dashed line) in $UFM$ phase (disentangled), then
it transits from the disentagled state to the entangled one in the
$ENQ$ phase, that is, an increase in the temperature makes the system
thermally entangled(the sudden emergence of quantum coherence) to
finally return to the non-entangled $UFM$ phase. In Fig.\ref{fig:diagrama}
(b), we display the phase diagram for both the entangled region and
the disentangled one as a function of the anisotropy parameter and
the threshold temperature, for a fixed value of the magnetic field
$h/J=2.2$ and $\Delta=1.3$. In this figure, the green dashed line
shows the transition going from the disentangled region to the entangled
one, and then back to the former(D-E-D). This transition is made possible
solely due to the re-entrance of the impurity bipartite entanglement.



\section{The thermal entangled teleportation}

\begin{figure}
\includegraphics[scale=0.44]{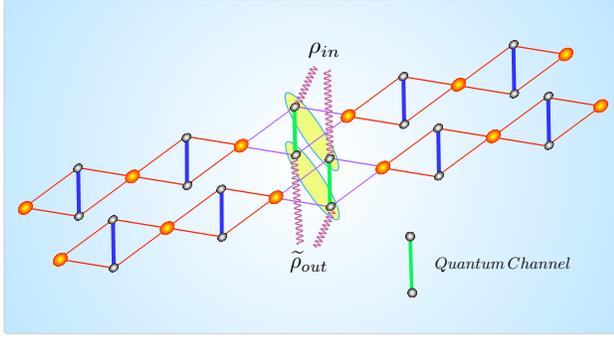}\caption{\label{fig:tele}The schematic representation for the teleportation
of the input state $\rho_{in}$, through a couple of independent impurity
Heisenberg dimers (green lines) in an Ising-$XXZ$ diamond chain.
The teleported output state is denoted by $\widetilde{\rho}_{out}$.}
\end{figure}

In this section, we implement teleportation throughout an entanglement
mixed state; it can be regarded as a general depolarizing channel
\cite{bo,peres} and we investigate the influence of the impurity
of the Ising-$XXZ$ diamond chain on the quantum teleportation. We
study the quantum teleportation via entangled states of a couple of
impurity Heisenberg dimer in the Ising-$XXZ$ diamond chain. We consider
two qubits in an arbitrary unknown state $|\psi_{in}\rangle$ as a
input, that is, 
\[
|\psi_{in}\rangle=\cos\left(\frac{\theta}{2}\right)|10\rangle+e^{i\phi}\sin\left(\frac{\theta}{2}\right)|01\rangle\;,
\]
where $0\leqslant\theta\leqslant\pi$ and $0\leqslant\phi\leqslant2\pi$.
Here, $\theta$ describes an arbitrary state and $\phi$ is the corresponding
phase of this state. In the density operator formalism, the concurrence
$\mathcal{C}_{in}$ of the input state, $\rho_{in}=|\psi_{in}\rangle\langle\psi_{in}|$,
can be written as 
\[
\mathcal{C}_{in}=2|e^{i\phi}\sin\left(\frac{\theta}{2}\right)\cos\left(\frac{\theta}{2}\right)|=|\sin(\theta)|\;.
\]
When the quantum state $\rho_{in}$, which is depicted in Fig. \ref{fig:tele},
is teleported via the mixed channel $\widetilde{\rho}_{ch}$, the
output replica state $\widetilde{\rho}_{out}$ can be obtained by
applying a joint measurement and the local unitary transformation
to the input state $\rho_{in}$\cite{bo,peres}, 
\[
\widetilde{\rho}_{out}=\sum_{i,j=\left\{ 0,x,y,z\right\} }p_{i}p_{j}\left(\sigma_{i}\otimes\sigma_{j}\right)\rho_{in}\left(\sigma_{i}\otimes\sigma_{j}\right)\;,
\]
in which $p_{i}=tr\left[E^{i}\widetilde{\rho}_{ch}\right]$, $E^{0}=|\Psi^{-}\rangle\langle\Psi^{-}|$,
$E^{1}=|\Phi^{-}\rangle\langle\Phi^{-}|$, $E^{2}=|\Phi^{+}\rangle\langle\Phi^{+}|$
and $E^{3}=|\Psi^{+}\rangle\langle\Psi^{+}|$, where $|\Phi^{\pm}\rangle=\frac{1}{\sqrt{2}}\left(|00\rangle\pm|11\rangle\right)$
and $|\Psi^{\pm}\rangle=\frac{1}{\sqrt{2}}\left(|01\rangle\pm|10\rangle\right)$
are Bell states. Here, we consider the density operator channel as
$\widetilde{\rho}_{ch}\equiv\widetilde{\rho}(T)$. 

\begin{figure}
\includegraphics[scale=0.4]{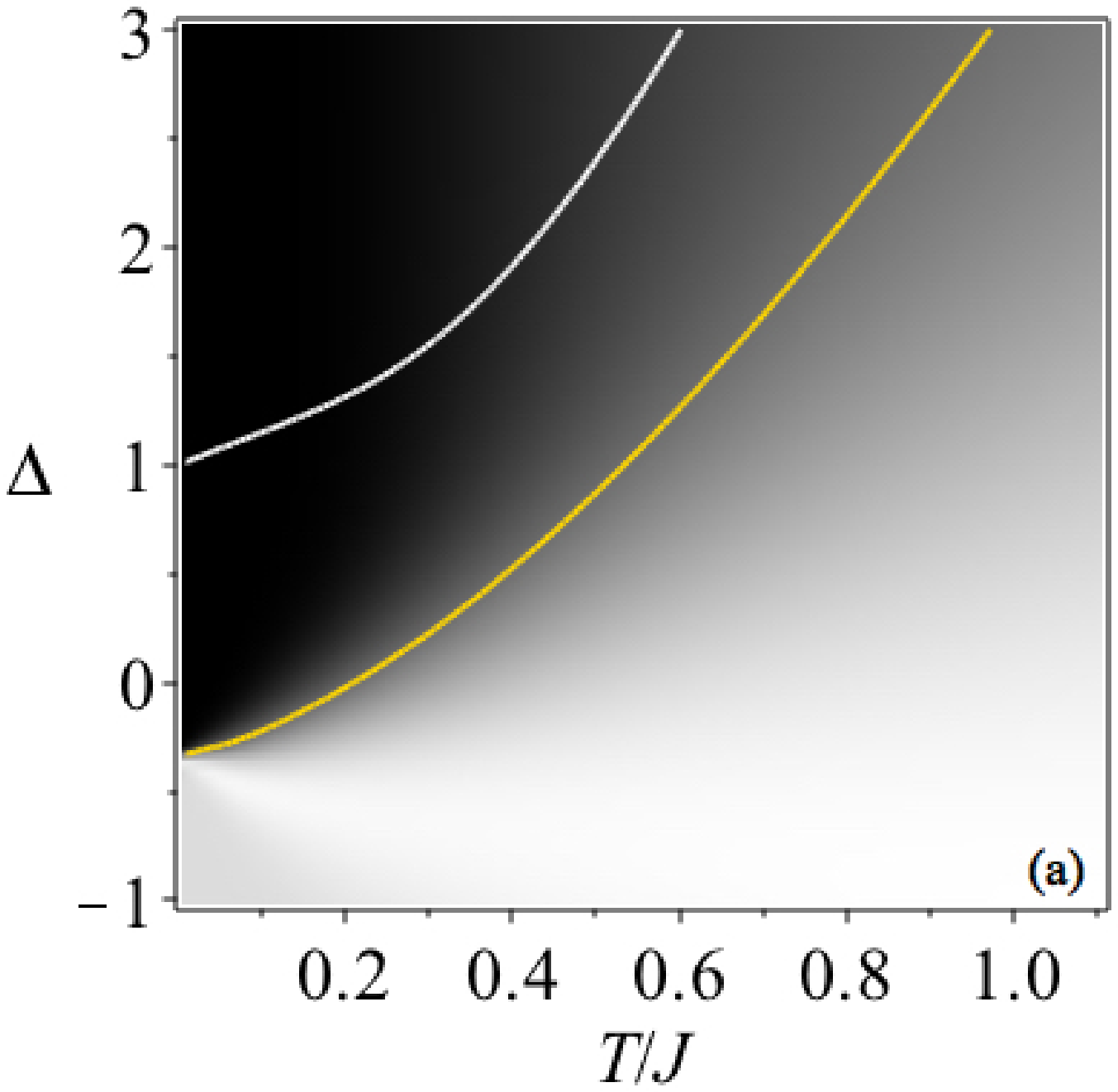}

\includegraphics[scale=0.4]{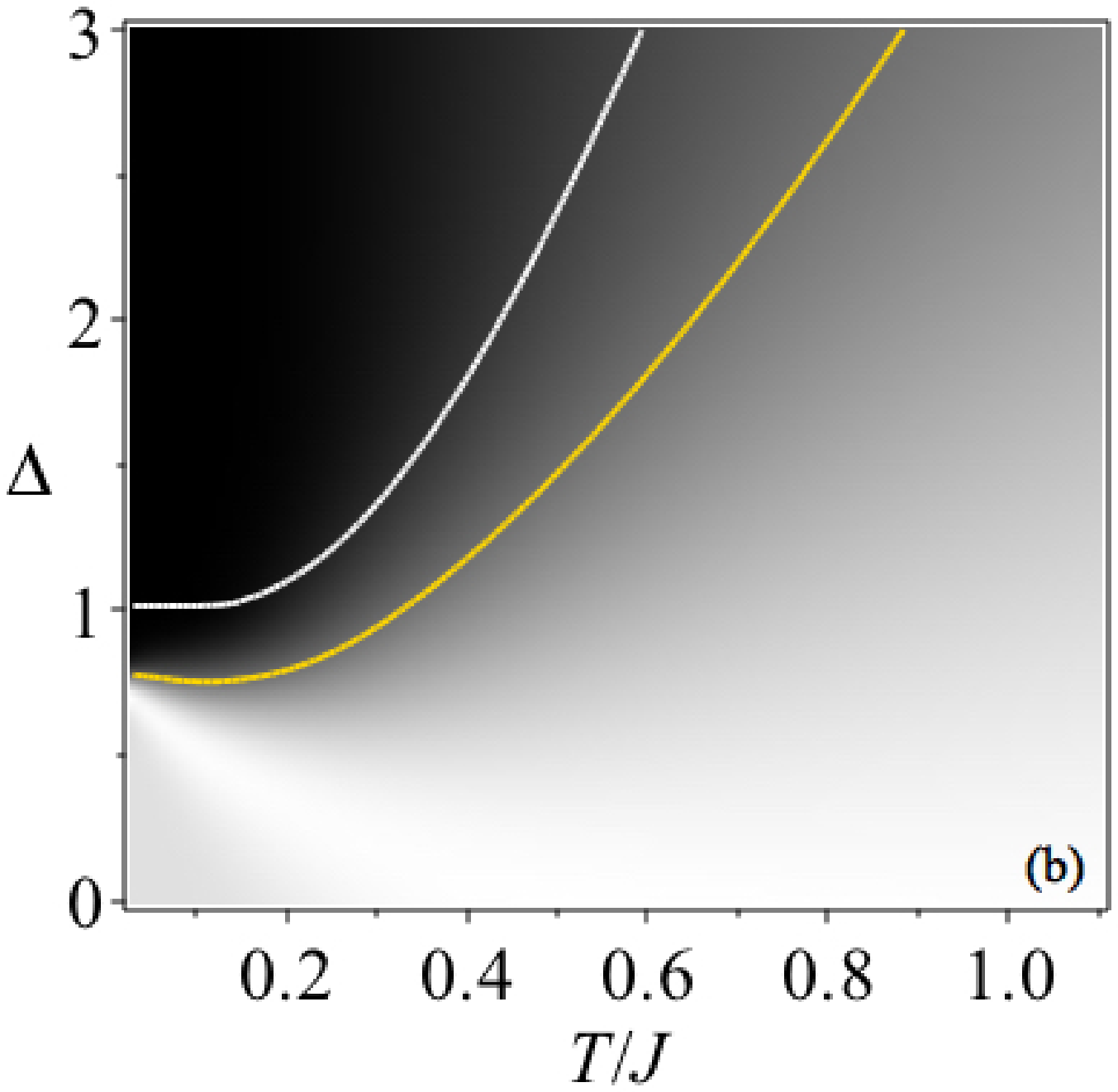}

\caption{\label{DFa}(Color online) The density plot of the average fidelity
$F_{A}$ as a function of $\Delta$ versus $T/J$. (a) $h/J=0$. (b)
$h/J=1.0$. In these figures, the yellow solid curve is the contour
for $F_{A}=2/3$ in the model with impurity and white one corresponds
to the model without it. The black(white) region corresponds to $F_{A}=1(0)$
and the gray regions indicate an average fidelity, $0<F_{A}<1$. }
\end{figure}

The output density operator $\widetilde{\rho}_{out}$ is described
by 
\begin{equation}
\widetilde{\rho}_{out}=\left[\begin{array}{cccc}
c & 0 & 0 & 0\\
0 & f & \Xi & 0\\
0 & \Xi & g & 0\\
0 & 0 & 0 & c
\end{array}\right]\;.\label{eq:rho-out}
\end{equation}
The elements of the operators can be expressed as 
\begin{flushleft}
\[
\begin{array}{cl}
c= & 2\widetilde{\rho}_{2,2}\left(\widetilde{\rho}_{1,1}+\widetilde{\rho}_{4,4}\right),\\
f= & \left(\widetilde{\rho}_{1,1}+\widetilde{\rho}_{4,4}\right)^{2}\cos^{2}\left(\frac{\theta}{2}\right)+4\widetilde{\rho}_{2,2}^{\,2}\sin^{2}\left(\frac{\theta}{2}\right),\\
g= & 4\widetilde{\rho}_{2,2}^{\,2}\cos^{2}\left(\frac{\theta}{2}\right)+\left(\widetilde{\rho}_{1,1}+\widetilde{\rho}_{4,4}\right)^{2}\sin^{2}\left(\frac{\theta}{2}\right),\\
\Xi= & 2e^{i\phi}\widetilde{\rho}_{2,3}^{\,2}\sin\theta\;.
\end{array}
\]
\par\end{flushleft}

By using the Eq. (\ref{eq:rho-out}) in the definition of concurrence,
Eq. (\ref{eq:Concurrence}), we obtain the output concurrence $\mathcal{C}_{out}(\widetilde{\rho})$
which is given by 
\[
\mathcal{C}_{out}(\widetilde{\rho})=2max\left\{ 2\widetilde{\rho}_{2,3}^{\,2}\mathcal{C}_{in}-2|\widetilde{\rho}_{2,2}||\widetilde{\rho}_{1,1}-\widetilde{\rho}_{4,4}|,0\right\} .
\]
More recently, the teleportation of the same entangled state was studied
in this model without any impurities \cite{moi-1}. 

\section{The Fidelity of entanglement teleportation}


In this section, we mainly focus on how much entanglement is teleported.
The fidelity between $\rho_{in}$ and $\rho_{out}$ characterizes
the quality of the teleported state. The fidelity is defined by \cite{Joz,shu}
\[
F=\langle\psi_{in}|\rho_{out}|\psi_{in}\rangle\;.
\]
After some algebra, one finds 
\begin{equation}
F=\frac{\sin^{2}\theta}{2}\left[\left(\widetilde{\rho}_{1,1}+\widetilde{\rho}_{4,4}\right)^{2}+4\widetilde{\rho}_{2,3}^{\,2}-4\widetilde{\rho}_{2,2}^{\,2}\right]+4\widetilde{\rho}_{2,2}^{\,2}.\label{eq:fidelity}
\end{equation}
The average fidelity $F_{A}$ of teleportation can be formulated as
\[
F_{A}=\frac{1}{4\pi}\intop_{0}^{2\pi}d\phi\intop_{0}^{\pi}F\sin\theta d\theta\;.
\]
According to the Eq.(\ref{eq:fidelity}), one can get the analytic
expression for $F_{A}$ as follows, 
\begin{equation}
F_{A}=\frac{1}{3}\left[\left(\widetilde{\rho}_{1,1}+\widetilde{\rho}_{4,4}\right)^{2}+4\widetilde{\rho}_{2,3}^{\,2}-4\widetilde{\rho}_{2,2}^{\,2}\right]+4\widetilde{\rho}_{2,2}^{\,2}\;.\label{eq:fidelityaverage}
\end{equation}
\begin{figure}
\includegraphics[scale=0.4]{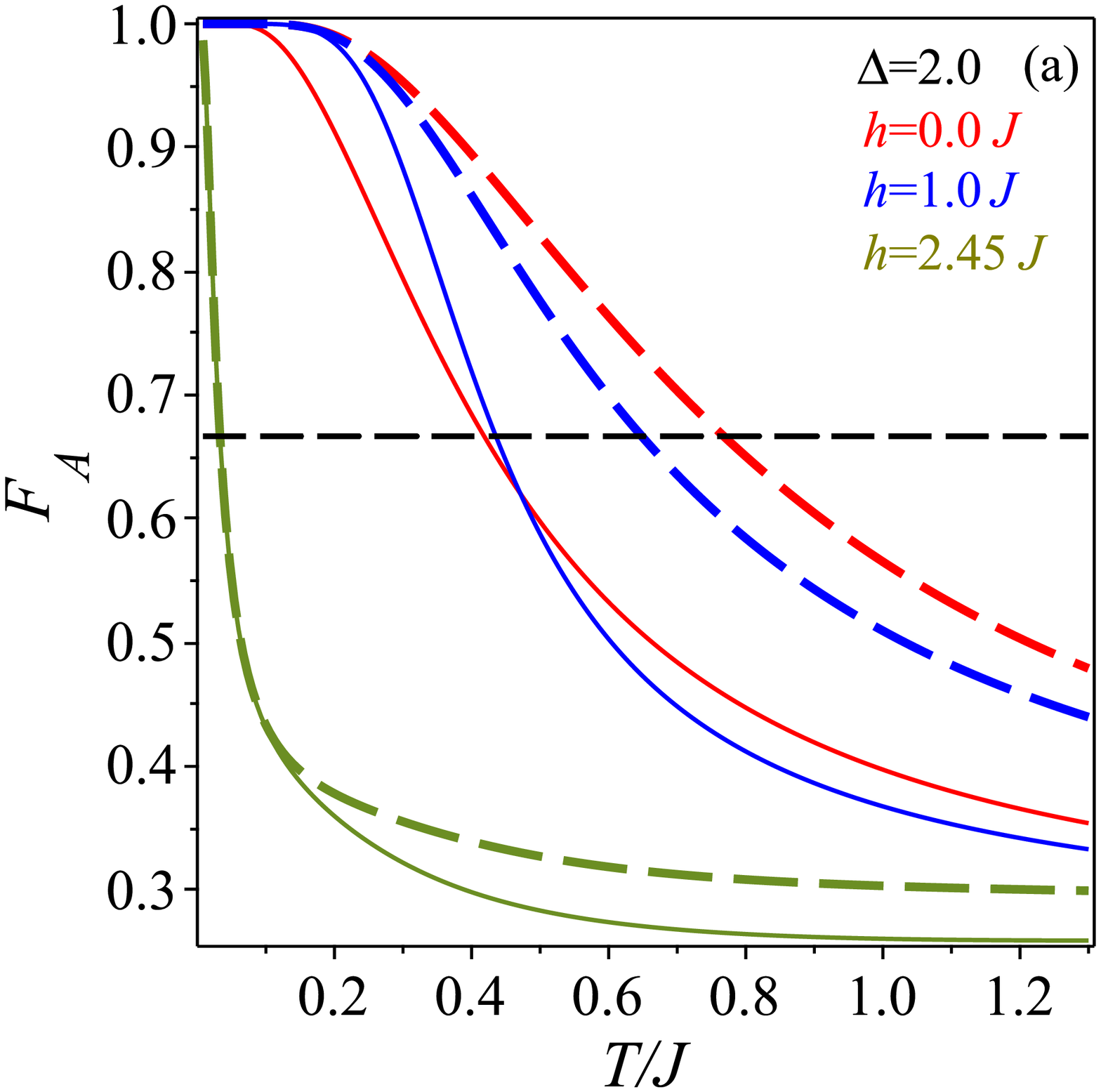}

\includegraphics[scale=0.4]{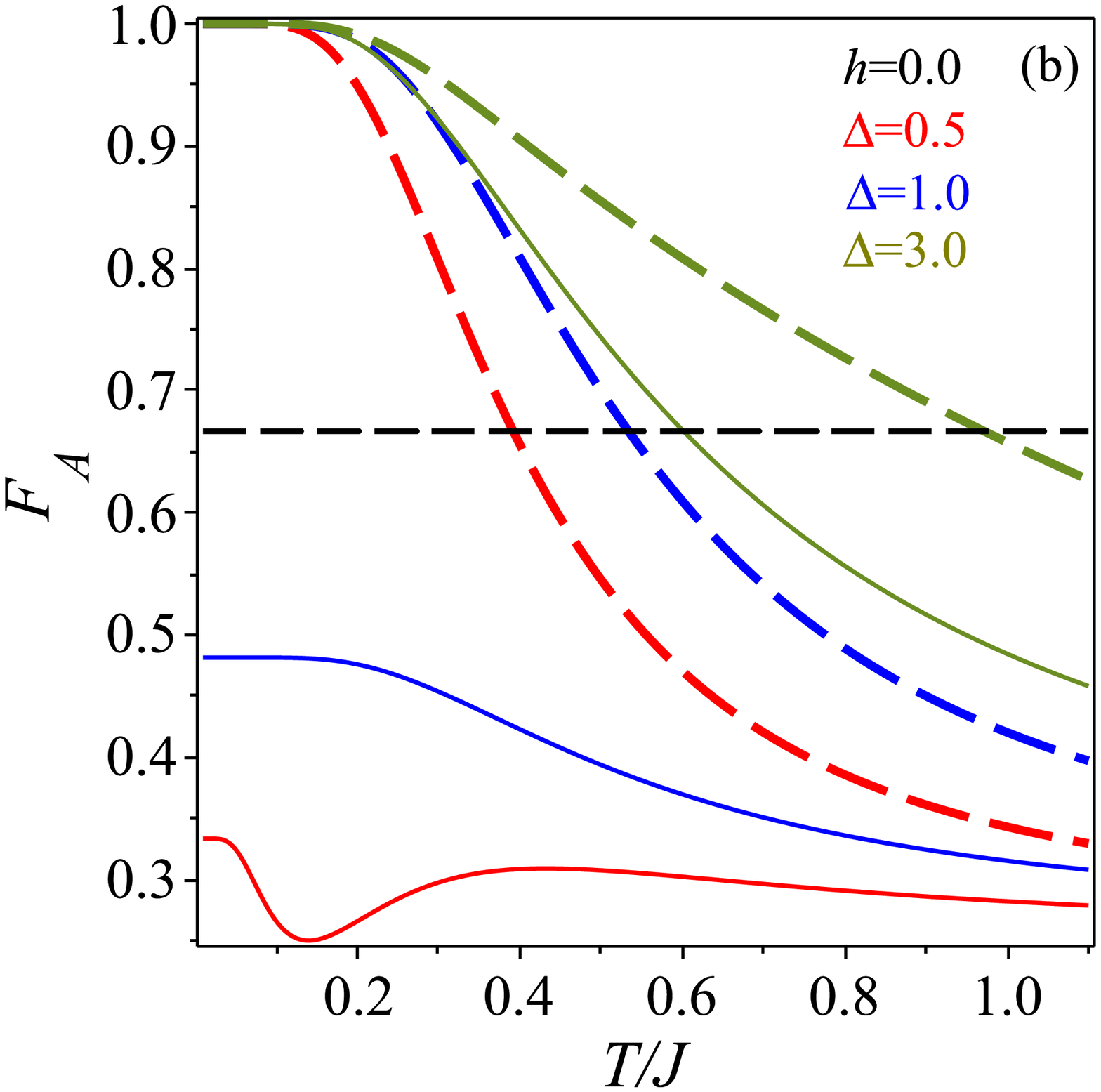}

\caption{\label{FavsT}(Color online) The average fidelity of the teleportation
$F_{A}$ is plotted against the temperature $T/J$. We consider $\alpha=0.0$,
$\gamma=0.8$, $\eta=-0.8$ and $J_{1}/J=1.0$. (a) $\Delta=2.0$.
(b) $h/J=0$. Horizontal dashed line indicates the $2/3$ constant
line.}
\end{figure}

The average fidelity $F_{A}$ is dependent on the quantum channel
parameters. In order to transmit the input state $|\psi_{in}\rangle$
with better fidelity than any classical communication protocol, $F_{A}$
must be greater than $2/3$. Taking into account the effects of an
impurity on the fidelity of entanglement teleportation, we compare
it with that given in the original model\cite{moi-1}. In Fig. \ref{DFa},
we illustrate the density plot of the average fidelity $F_{A}$ as
a function of $T/J$ and $\Delta$, for the two fixed values of the
magnetic field $h/J=0$ and $h/J=1.0$, respectively. The black region
corresponds to the maximum average fidelity ($F_{A}=1$), while that
the white one corresponds to $F_{A}=0$. In addition, the white (yellow)
curve is used to represent the surrounding. The dark region ($F_{A}>2/3$)
surrounded by curve white(yellow) indicates where the quantum teleportation
will become successful, whereas the outside means that the quantum
teleportation fails to be observed in the without(with) impurity case,
respectively. In the Fig. \ref{DFa}(a), it is depicted the density
plot of $F_{A}$ versus $\Delta$ and $T/J$, for the null magnetic
field. As can be observe in it, a wide region where the teleportation
of information is successful beyond the allowed region (limited for
white curve) in the impurity free case. This means that the introduction
of impurities allow the efficiency enhancement of the quantum teleportation.
It is also observed that the quantum teleportation of the isotropic
model ($\Delta=1$) is only possible at $T=0$, whereas in the presence
of impurity it is possible for $T\geqslant0$. In Fig. \ref{DFa}
(b), we have the density plot $F_{A}$ as a function of $\Delta$
and $T/J$, now for magnetic field $h/J=1.0$. In it, we also observe
an enhancement in the efficiency of the average fidelity due to the
inserted impurity. This increase is indicated by the region contorted
by both the white and yellow curves. In addition, in order to understand
the effects of the magnetic field $h/J$ and the anisotropic parameters
$\Delta$ on the average fidelity, in Fig. \ref{FavsT}, we plot $F_{A}$
as a function of the temperature $T/J$ under the conditions $\Delta=2.0$
and \index{}$h/J=0$, respectively. In Fig. \ref{FavsT}(a), we fixed
the anisotropic parameter as $\Delta=2.0$. From it, it is easy to
see that they can enhance the average fidelity for weak magnetic fields.
For higher magnetic fields and low temperatures the effect of the
impurity on the teleportation of information does not occurs. On the
other hand, in Fig. \ref{FavsT} (b), we fixed $h/J=0$ and we notice
that for either $\Delta=0.5$ or $\Delta=1.0$, the average fidelity
$F_{A}$ remains below $2/3$, making it impossible to the existence
of the quantum teleportation of the information. However, when we
consider the inclusion of impurity, we have a dramatic increase of
such a quantity and, at low temperatures, it reaches its maximum and
soon afterwards decreases monotonically as soon as the temperature
increases. Taking into account the strong anisotropy ($\Delta=3.0$),
the average fidelity is the same as that without impurity in low temperatures,
but by increasing the temperature, we observe a clear advantage of
the our model with impurity over the case without it. These results
show that we can get a significant enhancement of the fidelity by
the inclusion of impurities in the structure of the model.

\begin{figure}
\includegraphics[scale=0.4]{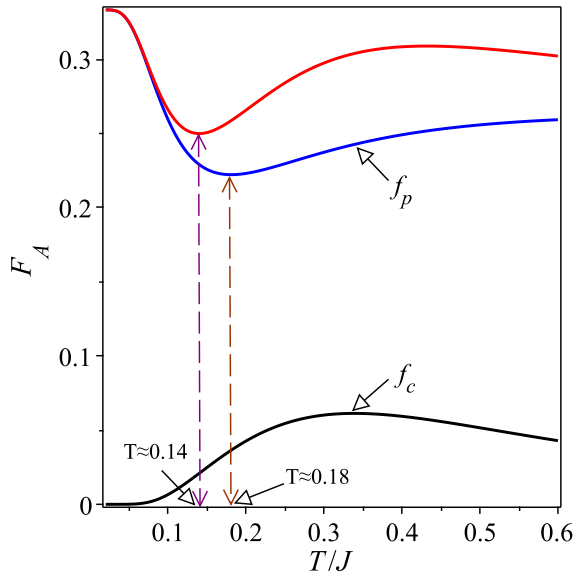}

\caption{\label{FavsT1}(Color online) The average fidelity of the teleportation
$F_{A}$, $f_{p}$, $f_{c}$ as a function of the temperature $T/J$
for the original model ($\alpha=0$, $\gamma=0$, $\eta=0$) with
$J_{1}/J=1.0$, $h=0$ and $\Delta=0.5$. Magenta vertical dashed
line indicates the minimum of the average fidelity and the brown vertical
dashed line indicate the minimum of the term $f_{p}$.}
\end{figure}

Finally, in the Fig. \ref{FavsT}(b), the average fidelity exhibits
intriguing non-monotonous temperature dependence for $\Delta=0.5$
in the classical communication region (red line). In order to understand
this behavior of the average fidelity, we rewrite Eq.(\ref{eq:fidelityaverage})
as $F_{A}=f_{p}+f_{c}$, where 
\[
\begin{array}{cl}
f_{p}= & \frac{1}{3}\left[\left(\rho_{1,1}+\rho_{4,4}\right)^{2}+8\rho_{2,2}^{\,2}\right],\\
f_{c}= & \frac{4}{3}\rho_{2,3}^{\,2}.\label{eq:fidelity-I}
\end{array}
\]
The term $f_{p}$ is a function of the elements of the population
of the density operator, while the term $f_{c}$ is a function of
the quantum coherence one. In Fig. \ref{FavsT1} , we observed the
term $f_{p}$ decays to a minimum at $T/J\approx0.18$, whereas the
term $f_{c}$ is initially null from $T=0$ to $T\approx0.05$, then
it increases the quantum coherence $f_{c}$ at a rate greater than
decay of therm $f_{p}$. This way, the average fidelity reaches the
minimum at $T\approx0.14$, increases until reaching a maximum and
then decreases monotonically. The curves of the average fidelity show
similar behavior in $0<\Delta<1$.




\section{Concluding remarks}

In summary, we have investigated the effects due to the inclusion
of an impurity plaquette on the spin-1/2 Ising-$XXZ$ diamond chain.
The concurrence and $l_{1}$-norm of the coherence is chosen as the
measurement of the thermal quantum correlation and they are obtained
by means of the transfer-matrix approach. We found that such a inclusion
in the system can induces a significant enhancement on the thermal
entanglement and on the quantum coherence. It is found that the concurrence
is more robust for low magnetic field. For strong magnetic fields,
we observed a sudden birth of the entanglement. Similarly, this behavior
also appeared in the quantum coherence. We have also discussed the
teleportation of the two-qubits in an arbitrary state through a couple
of quantum channel composed by impurity Heisenberg dimers in an infinite
Ising-$XXZ$ diamond chain. We observe how the teleportation of information
is successful beyond the allowed region in the impurity free case.
So, we showed that the average fidelity of teleportation could be
enhanced for some suitable impurity parameters. The influence of the
impurity is more evident in the average fidelity when we consider
the null magnetic field. We saw that it is possible to teleport information
in a wide range of anisotropic models. This is impossible in the model
without any impurity. As a final word, we state that considerable
enhancement of the teleportation can be achieved by tuning the strength
of the impurity parameters. This can be used locally to control the
quantum resources and the quantum teleportation of the information,
unlike the original model where it is globally done.

\section*{Acknowledgment}

M. Rojas and C. Filgueiras thank CNPq, Capes and FAPEMIG for partial
financial support. M. Freitas acknowledges support from Capes.


\begin{thebibliography}{10}
\bibitem{st} A. Streltsov, H. Kampermann, S. Wolk, M. Gessner, D.
Bru$\mathfrak{\beta}$, New J. Phys. \textbf{20}, 053058 (2018).

\bibitem{st-1} A. Streltsov, G. Adesso, M. B. Plenio, Rev. Mod. Phys.
\textbf{89}, 041003 (2017).

\bibitem{bra} C. H. Bennett, G. Brassard, C. Crepeau, C. Jozsa, A.
Peres, W. K. Wootters, Phys. Rev. Lett. \textbf{70}, 1895 (1993).

\bibitem{Bene} C. H. Bennett and D. P. Di Vincenzo, Nature \textbf{404,}
247 (2000).

\bibitem{amico} L. Amico, R. Fazio, A. Osterloh, V. Vedral, Rev.
Mod. Phys. \textbf{80,} 517 (2008).

\bibitem{rada} C. Radhakrishnan, M. Parthasarathy, S. Jambulingam,
T. Byrnes, Sci. Rep. \textbf{7,} 13865 (2017).

\bibitem{fan} G. Karpat, B. Çakmak, F. Fanchini, Phys. Rev. B \textbf{90,}
104431 (2014).

\bibitem{wei} W. Wu, J. Xu, Phys. Lett. A \textbf{381,} 239 (2017).

\bibitem{adesso} A. Streltsov, U. Singh, H. S. Dhar, M. N. Bera,
G. Adesso, Phys. Rev. Lett. \textbf{115,} 020403 (2015).

\bibitem{kam} G. L. Kamta, A. F. Starace, Phys. Rev. Lett. \textbf{88,}
107901 (2002); M. C. Amesen, S. Bose, V. Vedral, Phys. Rev. Lett.
\textbf{87,} 017901 (2001); X. G. Wang, Phys. Rev. A, \textbf{64,}
012313 (2001); J. Maziero, H. C. Guzman, L. C. Céleri, M. S. Sarandy,
R. M. Serra, Phys. Rev. A \textbf{82,} 012106 (2010); B. Çakmak, G.
Karpat, F. F. Fanchini, Entropy \textbf{17,} 790 (2015); J. Maziero,
H. Guzman, L. Céleri, M. Sarandy, R. Serra, Phys. Rev A \textbf{82,}
012106 (2010).

\bibitem{yeo} Y. Yeo, Phys Rev. A, \textbf{66,} 062312 (2002); G.
F. Zhang, Phys. Rev. A \textbf{75,} 034304 (2007).

\bibitem{kiku}H. Kikuchi, Y. Fujii, M. Chiba, S. Mitsudo, T. Idehara,
Physica B\textbf{ 329,} 967 (2003).

\bibitem{strec} M. Jascur, J. Strecka, J. Magn. Magn. Matter. \textbf{272,}
984 (2004); O. Rojas, S. M. de Souza, V. Ohanyan, M. Khurshudyan,
Phys. Rev. B \textbf{83,} 094430 (2011).

\bibitem{rojas} O. Rojas, S. M. de Souza, V. Ohanyan, M. Khurshudyan,
Phys. Rev. B\textbf{ 83,} 094430 (2011); L. Gálisová, Phys. Status
Solidi B\textbf{ 250,} 187 (2013); O. Rojas, S. M. de Souza, Phys.
Lett. A\textbf{ 375,} 1295 (2011).

\bibitem{moi} O. Rojas, M. Rojas, N. S. Ananikian, S. M. de Souza,
Phys. Rev. A \textbf{86,} 042330 (2012).

\bibitem{cheng} W. W. Cheng, X. Y. Wang, Y. B. Sheng, L. Y. Gong,
S. M. Yhao, J. M. Liu, Sci. Rep. \textbf{7,} 42360 (2017).

\bibitem{rojas-1} J. Torrico, M. Rojas, S. M. de Souza, O. Rojas,
N. S. Ananikian, Europhys. Lett. \textbf{108,} 50007 (2014); J. Torrico,
M. Rojas, M. S. S. Pereira, J. Strecka, M. L. Lyra, Phys. Rev. B \textbf{93,}
014428 (2016).

\bibitem{rojas-2} O. Rojas, M. Rojas, S. M. de Souza, J. Torrico,
J. Strecka, M. L. Lyra, Physica A \textbf{486,} 367 (2017).

\bibitem{moi-1} M. Rojas, S. M. de Souza, Onofre Rojas, Ann. Phys.
\textbf{377,} 506 (2017).

\bibitem{falk} H. Falk, Phys. Rev. \textbf{151,} 304 (1966); J. Stolze,
M. Vogel, Phys. Rev. B \textbf{61,} 4026 (2000).

\bibitem{fuku} T. Fukuhara, A. Kantian, M. Endres, M. Cheneau, P.
Schau$\beta$, S. Hild, D. Bellem, U. Schollw$\ddot{o}$ck, T. Giamarchi,
C. Gross, I. Bloch, S. Kuhr, Nature Physics \textbf{9,} 235 (2013).

\bibitem{xuchu} X. Huang, T. Si, Z. Yang, Physica B \textbf{462,}
25 (2015); X. Xi, S. Hao, W. Chen, R. Yue, Phys. Lett. \textbf{297},
291 (2002); T. J. G. Apollaro, F. Plastina, L. Banchi, A. Cuccoli,
R. Vaia, P. Verrucchi, M. Paternostro, Phys. Rev. A \textbf{88,} 052336
(2013).

\bibitem{fu} H. Fu, A. I. Solomon, X. Wang, J. Phys. A: Math. Gen.
\textbf{35,} 4293 (2002); W. W. Cheng, Y. X. Huang, T. K. Liu, H.
Li, Physica E \textbf{39,} 150 (2007); S. Li, J. Xu, Phys Lett. A
\textbf{334,} 109 (2005).

\bibitem{rojas-3} I. M. Carvalho, O. Rojas, S. M. de Souza, M. Rojas,
Quantum Inf. Process. \textbf{18}, 134 (2019).

\bibitem{baxter} R. J. Baxter, Exactly Solved Models in Statistical
Mechanics. Academic, New York (1982).

\bibitem{hill} S. Hill, W. K. Wootters, Phys. Rev. Lett. \textbf{78,}
5022 (1997).

\bibitem{woo} W. K. Wootters, Phys. Rev. Lett. \textbf{80}, 2245
(1998).

\bibitem{baum} T. Baumgratz, M. Cramer, M. B. Plenio, Phys. Rev.
Lett. \textbf{113}, 140401 (2014).

\bibitem{bo} G. Bowen, S. Bose, Phys. Rev. Lett. \textbf{87}, 267901
(2001).

\bibitem{peres} A. Peres, Phys. Rev. Lett. \textbf{77}, 1413 (1996).

\bibitem{Joz} R. Jozsa, J. Mod. Opt. \textbf{41}, 2315 (1994).

\bibitem{shu} B. Shumacher, Phys. Rev. A \textbf{54}, 2614 (1996). 
\end{thebibliography}
\end{document}